\documentclass[journal]{IEEEtran}
\usepackage{graphicx}
\usepackage{multicol,lipsum,amssymb}
\usepackage{mathtools, cuted}
\usepackage{mathtools}
\usepackage{amsmath}
\usepackage{etoolbox}
\usepackage{hyperref}
\usepackage{lipsum}
\usepackage{csquotes}
\usepackage{stfloats}

\patchcmd{\thebibliography}{\advance\leftmargin\labelsep}{}{}{}

\ifCLASSINFOpdf
\else
\fi
\begin{document}

\title{On the Physical Layer Security of a Decode and Forward Based Mixed FSO/RF Cooperative System}

\author{Dipti R. Pattanayak,~\IEEEmembership{Student Member,~IEEE},
Vivek K. Dwivedi,~\IEEEmembership{Member,~IEEE,}\\
         Vikram Karwal,~\IEEEmembership{Senior Member,~IEEE}, Imran Shafique Ansari,~\IEEEmembership{Member,~IEEE}, Hongjiang Lei,~\IEEEmembership{Member,~IEEE} and Mohamed-Slim Alouini,~\IEEEmembership{ Fellow, IEEE}
}


\maketitle

\begin{abstract}
In this letter, the secrecy performance of a mixed free space optics (FSO) and radio frequency (RF) system is analyzed  from physical layer security (PHY) perspective. In this scenario, one or more eavesdroppers are trying to intercept the confidential signal in a mixed FSO/RF system. The faded FSO links are modeled by M{\'a}laga ($\mathcal{M} $) distribution and RF link is characterized by Nakagami-$m$ distribution. Exact closed form expressions for secrecy performance metrics such as secrecy outage probability and strictly positive secrecy capacity are derived and analyzed for the proposed system in terms of Fox's H-function. Furthermore, the asymptotic expressions for  these performance metrics are analyzed. Finally, all the results are verified by Monte-Carlo simulations.

\end{abstract}


\IEEEpeerreviewmaketitle
\section{Introduction}
\IEEEPARstart{P}{hysical} Layer Security and protection have progressively become issues of interest in wireless communication network [1-9]. 
 Wireless communication is more vulnerable to eavesdropping because of its broadcast nature. An enormous amount of classified and sensitive data e.g. budget information, client documents, military information, healthcare records, etc. are transmitted by means of wireless channel. The inherent openness of wireless medium permits anyone inside the scope to track the signal, which makes the data security one of the most critical and troublesome issues in wireless systems.
Traditionally the data security is handled by cryptographic approach (used in Application layer (layer 7), network layer (layer 3), and data-link layer (layer 2)) \cite{2}. Generally, cryptographic approach is based on the data encryption method. Because of advancement in breaking of encrypted algorithms, this method is always vulnerable for security. Subsequently, new security approaches are incorporated based on the information theory fundamentals i.e. focusing on the secrecy capacity of the channel.  This is termed as physical layer security (PHY) and it is handled in the physical layer [3-7]. PHY solution is basically achieved by using different methods such as code based, signalling based, and physical layer encryption based methods.

Many researchers put their effort to analyze the secrecy performance from RF perspective by analyzing the secrecy performance for various fading channels. The performance of a wiretap channel was first introduced by Wyner \cite{1} in which a source is trying to communicate with the destination in the presence of an external eavesdropper. Based on the nature of the propagation channel, researchers focused on various fading distributions to mitigate the piracy such as: generalized $K$ \cite{2}, $\alpha$-$\mu$ \cite{3}, generalized Gamma \cite{4} etc.
On the other side due to channel scattering, atmospheric turbulence, pointing deflection and optical beam divergence \cite{5}, there is a possibility of eavesdropping of confidential data. Therefore authors in \cite{6} pointed out the data eavesdropping due to optical beam divergence and turbulence effect. The analysis in \cite{6} also concludes that     
a eavesdropper can be able to overhear the data if it is very close to the legitimate node. Then in \cite{7}, it is investigated that the information can be overheard from outside the laser beam converging area through an non-line of sight (NLOS) scattering channel. In \cite{8} it is highlighted that the eavesdropping also possible, if the beam  is reflected by a small particle or blocked by a suitable obstacle. Then some efforts have been devoted to analyze the FSO link from PHY perspective such as: in \cite{8}, authors explore the PHY for a FSO channel characterized by  M{\'a}laga ($\mathcal{M} $) distribution and analysed the secrecy outage probability (SOP), strictly positive secrecy capacity (SPSC) and average secrecy capacity (ASC) as a performance metrics.

 Moreover, the research was extended for cooperative relaying based approach to achieve the secure communication \cite{VPoor}. Such relaying approach is mainly used in mixed FSO/RF environments to get the benefits of FSO to the end user level. Authors in \cite{9} considered a mixed RF/FSO (uplink) environment by analysing the performance metrics such as: SOP and ASC, where the RF link is modeled by Nakagami-$m$ distribution and FSO link is characterized by Gamma-Gamma (GG) distribution. During analysis the eavesdropper is considered to be located in th RF link, modelled by Nakagami-$m$ distribution. Furthermore, in \cite{10}, the outage probability, average symbol error rate, channel capacity and intercept probability for a mixed RF/FSO system were analysed from RF security perspective. Similarly in  \cite{11}, the PHY analysis is done for a hybrid satellite-FSO co-operative system  where the satellite link is modelled by the shadowed Rician distribution and FSO link is assumed to be Gamma-Gamma  fading distribution. Security analysis was further extended  for downlink RF/FSO system i.e. FSO/RF system in \cite{12}. The FSO/RF system has the advantage to serve multiple RF users with different data rates \cite{13}. However, reported works in RF/FSO and FSO/RF scenarios are limited to the secrecy performance analysis for  the RF  side eavesdroppers only.
 This motivates to explore the PHY performance for a FSO/RF system, where an optical eavesdropper at relay is trying to interrupt the communication. The main contributions of this work are:        
\begin{enumerate}
\item So far no literature on PHY analysis is available in the context of information overhearing from RF based eavesdropper as well as FSO based eavesdropper simultaneously in a mixed FSO/RF system. More specifically during analysis, we have considered the worst case scenario by assuming that the confidential data is intercepted in each hop of the system i.e. eavesdroppers are trying to intercept the signal from FSO as well as RF links simultaneously.

\item The closed form expressions for secrecy performance metrics  such as SOP and SPSC in wide range of fading severity (i.e. moderate, and strong regions) is derived  under the impact of pointing error and different FSO data detection methods.
\item The SOP and SPSC expressions are obtained in terms of Meijer's-function and bivariate Fox's H-function (i.e. EGBFHF).
\item Further, the derived closed form results are expressed asymptotically to validate the proposed work and finally all the results are validated through Monte-Carlo simulation.
\end{enumerate}

\section{System and channel models}
A mixed FSO/RF wireless scenario is considered in which the source (S) transmits its private data to the legitimate destination node (D) through a trusted relay (R). During transmission, two unauthorized receivers (E1 and E2) are trying to decode the data using a photo detector at the relay and a RF antenna at the destination respectively. In this model, the FSO Link-1 (i.e. S--R) and  FSO Link-2 (i.e. S--E1) are assumed to follow M{\'a}laga ($\mathcal{M} $) distribution. The relay  is connected to the destination through a RF link that is modelled by Nakagami-$m$ distribution. The channel between relay and eavesdropper (E2) is also assumed to follow the Nakagami-$m$ distribution. A decode and forward (DF) based relay is considered that decodes the incoming signal and retransmits to the destination. The relay has a photo detector to receive the optical signal from the source and has a RF antenna to retransmit the RF signal to the destination.

The probability density function (PDF) and the cumulative distribution function (CDF) of the instantaneous SNR (${\gamma_{Sk}}  ;  k\in\left\{ R,E_{1}\right\} $) for M{\'a}laga fading channels are expressed as \cite{14}
\begin{equation}\label{eq:1}
\small f_{\gamma_{Sk}}(\gamma)=\frac{A\xi^{2}}{2^{r}\gamma}\sum_{m=1}^{\beta}b_{m}G_{1,3}^{3,0}\left[B\left(\frac{\gamma}\mu_{r}\right)^{\frac{1}{r}}\Bigg|\begin{array}{c}
\xi^{2}+1\\
\xi^{2},\alpha,m
\end{array}\right],
\end{equation}

\begin{equation}\label{eq:2}
F_{\gamma_{Sk}}(\gamma)=\frac{A\xi^{2}}{2^{r}(2\pi)^{r-1}}\sum_{m=1}^{\beta}c_{m}G_{r+1,3r+1,}^{3r,1}\left[E\frac{\gamma}\mu_{r}\Bigg|\begin{array}{c}
1,K_{1}\\
K_{2},0
\end{array}\right],
\end{equation}
where $r=1$ and $r=2$ represents the average SNR for heterodyne (HD) and intensity modulation
with direct detection (IM/DD) techniques respectively with $\mu_{1}=\overline{\gamma}_{Sk}$ and $\mu_{2}=\frac{\xi^{2}\left(1+\xi^{2}\right)^{-2}\left(2+\xi^{2}\right)\left(g+\Omega'\right)}{\alpha^{-1}\left(1+\alpha\right)\left[2g\left(g+2\Omega'\right)+\Omega'^{2}\left(1+\frac{1}{\beta}\right)\right]}\overline{\gamma}_{Sk}$. $A=\frac{2\alpha^{\frac{\alpha}{2}}}{g^{1+\frac{\alpha}{2}}r\Gamma\left(\alpha\right)}\left(\frac{g\beta}{g\beta+\Omega^{'}}\right)^{\beta+\frac{\alpha}{2}}$, $B=\frac{\xi^{2}\alpha\beta\left(g+\Omega^{'}\right)}{\left(\xi^{2}+1\right)\left(g\beta+\Omega^{'}\right)}$, $b_{m}=a_{m}\left[\frac{\alpha\beta}{g\beta+\Omega^{'}}\right]^{-\frac{\alpha+m}{2}}$, $c_{m}=b_{m}r^{\alpha+m-1}$, $a_{m}=\left(\begin{array}{c}
\beta-1\\
m-1
\end{array}\right)\frac{\left(g\beta+\Omega^{'}\right)^{1-\frac{m}{2}}}{\left(m-1\right)!}\left(\frac{\Omega^{'}}{g}\right)^{m-1}\left(\frac{\alpha}{\beta}\right)^{\frac{m}{2}}$, $E=\frac{B^{r}}{r^{2r}}$. The parameter $\alpha$ is a positive number related to effective number of large scale cells of the scattering process, and $\beta$ shows the fading severity of the channel. From \cite{14}, $g=2b_{0}\left(1-\rho\right)$, denotes the average power of the
scattering component received by off-axis eddies, $2b_{0}$ is the average power of the total scattering components.  The parameter $\rho$ denotes amount of scattering power
coupled to the line-of-sight (LOS) component with $0\leq\rho\leq1$, $\Omega^{'}=\Omega+2b_{0}\rho+2\sqrt{2b_{0}\rho\Omega}cos\left(\phi_{A}-\phi_{B}\right)$ represents the average power from coherent contributions, $\Omega$ is the average power of the LOS component, and $\phi_{A}$ and $\phi_{B}$ are the deterministic phases of the LOS and the coupled-to-LOS scatter items, respectively \cite{14}. Parameter $r$ represents the scheme of detection at the photo receiver. $\xi$ is the measure of pointing deflection between transmitter and receiver. $\overline{\gamma}_{Sk}$ and $\overline{\gamma}_{RD}$ are the electrical and average SNRs for the respective links. $K_{1}=\left[\Delta\left(r,\xi^{2}+1\right)\right]$, $K_{2}=\left[\Delta\left(r,\xi^{2}\right),\Delta\left(r,\alpha\right),\Delta\left(r,m\right)\right]$ where $\Delta\left(x,y\right)=\frac{y}{x},\frac{y+1}{x},...,\frac{y+x-1}{x}$, $G_{p,q}^{m,n}\left[.\right]$ is the Meijer's G function [16, Eq.  (9.301)], and $\Gamma(.)$ is the Gamma function [16, Eq. (8.310)].
 
 The faded RF link is assumed to experience a Nakagami-$m$ distribution \cite{16}. The PDF and CDF for the instantaneous SNR (${\gamma_{Rj}}  ;  j\in\left\{ D,E_{2}\right\} $) can be written as follows
 \begin{equation}\label{eq:3}
f_{\gamma_{Rj}}\left(\gamma\right)=\left(\frac{m_{Rj}}{\overline{\gamma}_{Rj}}\right)^{m_{Rj}}\frac{\gamma^{\left(m_{Rj}-1\right)}}{\Gamma\left(m_{Rj}\right)}\exp\left(-\frac{m_{Rj}\gamma}{\overline{\gamma}_{Rj}}\right),
  \end{equation}

\begin{equation}\label{eq:4}
F_{\gamma_{Rj}}\left(\gamma\right)=1-\sum_{k=0}^{m-1}\frac{1}{k!}\left(\frac{m\Theta\gamma}{\overline{\gamma}_{RD}}\right)^{k}\exp\left(-\frac{m\Theta\gamma}{\overline{\gamma}_{RD}}\right).
\end{equation}

For a DF based relaying network, the end-to-end SNR can be written as $\gamma_{eq}=$ $\min$ $\left(\gamma_{SR,}\gamma_{RD}\right)$ \cite{16}.  From \cite{16}, the CDF for $\gamma_{eq}$ is obtained as
\begin{equation}\label{eq:5}
F_{\gamma_{eq}}\left(\gamma\right)=1-\Pr\left[\gamma_{SR}>\gamma\right]\Pr\left[\gamma_{RD}>\gamma\right].
\end{equation}
Using Eq.\eqref{eq:1} and Eq.\eqref{eq:3} in Eq.\eqref{eq:4} and utilizing [18, Eq. (07.34.21.0085.01)], the expression for end-to-end (S-R-D) CDF is obtained as 
\begin{multline}\label{eq:6}
F_{\gamma_{eq}}\left(\gamma\right)=1-\frac{A\xi^{2}2^{-r}}{(2\pi)^{r-1}}\exp\left(-\frac{m_{RD}\gamma}{\overline{\gamma}_{RD}}\right)\sum_{k'=0}^{(m_{RD}-1)}\sum_{m=1}^{\beta}\frac{b_{m}}{r^{1-\alpha-m}}\\\times\frac{\left(m_{RD}-1\right)}{k'!\Gamma\left(m_{RD}\right)}\left(\frac{m_{RD}\gamma}{\overline{\gamma}_{RD}}\right)^{k'}G_{r+1,3r+1,}^{3r+1,0}\left[\frac{B^{r}\gamma}{r^{2r}\mu_{r}}|\begin{array}{c}
1,K_{1}\\
K_{2},0
\end{array}\right].
\end{multline}

\section{ Eavesdropping Attack on FSO Link Only}
\subsection{Secrecy outage probability analysis}

SOP is one of the fundamental secrecy benchmarks, which shows the level of privacy in any communication system. It is defined as the probability that instantaneous secrecy capacity ($C_{s}$) falls below a target secrecy rate \cite{4} and it can be expressed as
\begin{equation}\label{eq:7}
P_{out}\left(r_{s}\right)=Pr\left\{ C_{s}\left(\gamma_{eq},\gamma_{SE}\right)\leq R_{s}\right\}, 
\end{equation}
where $R_s$ is the target secrecy rate. Referring to \cite{4}, the lower bound for the SOP is derived as 
\begin{equation}\label{eq:8}
SOP_{L}=\int_{0}^{\infty}F_{\gamma_{eq}}\left(\varTheta\gamma\right)f_{\gamma_{SE}}\left(\gamma\right)d\gamma.
\end{equation}
The expression of secrecy outage probability for FSO/RF system can be expressed as given in \eqref{eq:9}
  , where $\varTheta=e^{R_{s}}$,  $\upsilon_{1}=\frac{B^{r}}{\mu m_{RD}r^{2r}}$,  $\upsilon_{2}=\frac{B^{r}\overline{\gamma}_{RD}}{\mu\overline{\gamma}_{SE}m_{SE}\Theta}$,  and $\left[Q\right]_{i}=Q,Q,...,Q$, comprising $i$ terms.  $H_{p_{1},q_{1};p_{2},q_{2};p_{3},q_{3}}^{m_{1},n_{1};m_{2},n_{2};m_{3},n_{3}}\left[.\right]$ is the H-function of two variables, which can be efficiently implemented in MATLAB \cite{18}. 
\vspace{0.5pc}
\newline 
\emph{Proof}: See appendix A.
\newline

  \begin{figure*}[t]
$SOP_{L_{1}}=1-\frac{\left(m_{RD}-1\right)}{\Gamma\left(m_{RD}\right)}\left(\frac{\xi^{2}A}{2^{r}}\right)^{2}\frac{1}{(2\pi)^{r-1}}\sum_{k'=0}^{(m_{RD}-1)}\left(\frac{m_{RD}\varTheta}{\overline{\gamma}_{RD}}\right)^{k'}\frac{1}{k'!}\sum_{m_{1}=1}^{\beta_{1}}\sum_{m_{2}=1}^{\beta_{2}}$
\begin{equation}\label{eq:9}
\times r^{\alpha+m_{1}}b_{m_{1}}b_{m_{2}}H_{1,0;r+1,3r+1;1,3}^{0,1;3r+1,0;3,0}\\\left[\upsilon_{1},\upsilon_{2}|\begin{array}{c}
\left(1;1,1\right)\\
\left(-;-,-\right)
\end{array}|\begin{array}{c}
\left(K_{1},\left[1\right]_{k_{1}}\right),(1,1)\\
\left(K_{2},\left[1\right]_{k_{2}}\right),(0,1)
\end{array}|\begin{array}{c}
(\xi^{2}+1,r)\\
(\xi^{2},r),(\alpha,r),(m_{2},r)
\end{array}\right].
\end{equation}
\hrulefill
\end{figure*}

 The SOP in Eq. \eqref{eq:9} reduces to the SOP for Gamma-Gamma- Rayleigh (for m = 1) special case as 
\begin{equation}
SOP_{L_{1}}=1-\frac{A^{2}r^{\alpha+\beta}}{(2\pi)^{r-1}}H_{1,0;r+1,3r+1;1,3}^{1,0;3r+1,0;3,0}\nonumber
\end{equation}
 \begin{footnotesize}
 \begin{align}\label{eq:10}
\times\left[\varXi_{1},\varXi_{2}\Bigg|\begin{array}{c}
\left(1;1,1\right)\\
\left(-;-,-\right)
\end{array}\Bigg|\begin{array}{c}
\left(K_{1}^{'},\left[1\right]_{k_{1}}\right),(1,1)\\
\left(K_{2}^{'},\left[1\right]_{k_{2}}\right),(0,1)
\end{array}\Bigg|\begin{array}{c}
(\xi^{2}+1,r)\\
(\xi^{2},r),(\alpha,r),(\beta,r)
\end{array}\right],
  \end{align}
  \end{footnotesize}
where $A=\frac{\xi^{2}}{r\Gamma\left(\alpha\right)\Gamma\left(\beta\right)}$, $\varXi_{1}=\frac{\left(h\alpha\beta\right)^{r}\overline{\gamma}_{RD}}{\overline{\gamma}_{SR}r^{2r}}$, $\varXi_{2}=\frac{\left(h\alpha\beta\right)^{r}\overline{\gamma}_{RD}}{\overline{\gamma}_{SE}\Theta}$, $K_{1}^{'}=\left[\Delta\left(r,\xi^{2}+1\right)\right]$ and $K_{2}^{'}=\left[\Delta\left(r,\xi^{2}\right),\Delta\left(r,\alpha\right),\Delta\left(r,\beta\right)\right]$.

At high SNR condition, $\overline{\gamma}\rightarrow\infty$ i.e. $Z\rightarrow\infty$, the Meijer's G function expansion [15, Eq.41] is expressed as
\begin{multline}\label{eq:11}
\lim_{Z\rightarrow\infty}G_{p,q}^{m,n}\left[Z\Bigg|\begin{array}{c}
a_{1},...,a_{p}\\
b_{1},...,b_{q}
\end{array}\right]\\=\sum_{k=1}^{n}\frac{\prod_{l=1;l\neq k}^{n}\Gamma\left(a_{k}-a_{l}\right)\prod_{l=1}^{m}\Gamma\left(1+b_{l}-a_{k}\right)}{\prod_{l=n+1}^{p}\Gamma\left(1+a_{l}-a_{k}\right)\prod_{l=m+1}^{q}\Gamma\left(a_{k}-b_{l}\right)}Z^{a_{k}-1}.
\end{multline} 

Upon substituting Eq. \eqref{eq:1} and Eq. \eqref{eq:6} in Eq. \eqref{eq:8} and applying [18, Eq. (07.34.21.0013.01)] and Eq. \eqref{eq:11}, the asymptotic expression of Eq. \eqref{eq:8} can be expressed as  

\begin{multline}\label{eq:12}
SOP_{L_{1}}^{\infty}\underset{\overline{\gamma}>>1}{\approxeq}\sum_{m_{1}=1}^{\beta_{1}}\sum_{m_{2}=1}^{\beta_{2}}\left(\frac{\xi^{2}A}{2^{r}(2\pi)^{r-1}}\right)^{2}r^{\alpha+m_{2}-1}c_{m_{1}}b_{m_{2}}\\\times\sum_{k=1}^{3r}\left(\frac{\overline{\gamma}_{SR}}{\overline{\gamma}_{SE}\varTheta}\right)^{K_{3,k}-1}\frac{\prod_{l=1;l\neq k}^{3r}\Gamma\left(K_{3,k}-K_{3,l}\right)}{\prod_{l=3r+1}^{4r+1}\Gamma\left(1+K_{3,l}-K_{4.k}\right)}\\\times\frac{\prod_{l=1}^{3r+1}\Gamma\left(1+K_{4,l}-K_{3.k}\right)}{\prod_{l=3r+2}^{4r+1}\Gamma\left(K_{3,k}-K_{4.l}\right)},
\end{multline}
where $K_{i,j}$ denotes the $j_{th}$ term of $K_i$ and $i\in\left\{ 3,4\right\} $ with  $K_{3}=[1-\triangle\left(r,\xi^{2}\right),1-\triangle\left(r,\alpha\right),1-\triangle\left(r,m_{1}\right),1,\triangle\left(r,1+\xi^{2}\right)]$ and $K_{4}=[\triangle\left(r,\xi^{2}\right),\triangle\left(r,\alpha\right),\triangle\left(r,m_{2}\right),0,1-\triangle\left(r,2-\xi^{2}\right)]$. 
\subsection{Strictly positive secrecy capacity analysis \textup{(SPSC)}}
SPSC shows the probability of existence of secrecy capacity. It can be formulated as follows \cite{2}
\begin{equation}
SPSC_{1}=Pr\left\{ C_{s}\left(\gamma_{eq},\gamma_{SE}\right)>0\right\}\nonumber
\end{equation}
\begin{equation}\label{eq:13}
=1-SOP_{L_{1}}|_{R_{S}=0}.
\end{equation}
Upon substituting $R_{S}=0$ in \eqref{eq:9} and then substituting in \eqref{eq:13}, the SPSC can be expressed as Eq. \eqref{eq:14}, where $\upsilon_{2}^{'}=\frac{B^{r}\overline{\gamma}_{RD}}{\mu\overline{\gamma}_{SE}m_{SE}}$. The SPSC in Eq. \eqref{eq:14} can be approximated at high SNR as 

\begin{figure*}
$SPSC_{1}=\frac{\left(m_{RD}-1\right)}{\Gamma\left(m_{RD}\right)}\left(\frac{\xi^{2}A}{2^{r}}\right)^{2}\frac{1}{(2\pi)^{r-1}}\sum_{k'=0}^{(m_{RD}-1)}\left(\frac{m_{RD}\varTheta}{\overline{\gamma}_{RD}}\right)^{k'}\frac{1}{k'!}\sum_{m_{1}=1}^{\beta_{1}}\sum_{m_{2}=1}^{\beta_{2}}$
\begin{equation}\label{eq:14}
\times r^{\alpha+m_{1}}b_{m_{1}}b_{m_{2}}H_{1,0;r+1,3r+1;1,3}^{0,1;3r+1,0;3,0}\left[\upsilon_{1},\upsilon_{2}^{'}|\begin{array}{c}
\left(1;1,1\right)\\
\left(-;-,-\right)
\end{array}|\begin{array}{c}
\left(K_{1},\left[1\right]_{k_{1}}\right),(1,1)\\
\left(K_{2},\left[1\right]_{k_{2}}\right),(0,1)
\end{array}|\begin{array}{c}
(\xi^{2}+1,r)\\
(\xi^{2},r),(\alpha,r),(m_{2},r)
\end{array}\right].
\end{equation}
\hrulefill
\end{figure*}

\begin{equation}\label{eq:15}
SPSC_{1}^{\infty}\underset{\overline{\gamma}>>1}{\approxeq}1-SOP_{L}^{\infty}|_{R_{S}=0}.
\end{equation}

Correspondingly, the asymptotic expansion for SPSC is obtained as given in Eq. \eqref{eq:16}.

\begin{figure*}
$SPSC_{1}^{\infty}\underset{\overline{\gamma}>>1}{\approxeq}\sum_{m_{1}=1}^{\beta_{1}}\sum_{m_{2}=1}^{\beta_{2}}\left(\frac{\xi^{2}A}{2^{r}(2\pi)^{r-1}}\right)^{2}r^{\alpha+m_{2}-1}c_{m_{1}}b_{m_{2}}$
\begin{equation}\label{eq:16}
\times \sum_{k=1}^{3r}\left(\frac{\overline{\gamma}}{\overline{\gamma}_{SE}}\right)^{K_{3,k}-1}\times\frac{\prod_{l=1;l\neq k}^{3r}\Gamma\left(K_{3,k}-K_{3,l}\right)}{\prod_{l=3r+1}^{4r+1}\Gamma\left(1+K_{3,l}-K_{4.k}\right)}\times\frac{\prod_{l=1}^{3r+1}\Gamma\left(1+K_{4,l}-K_{3.k}\right)}{\prod_{l=3r+2}^{4r+1}\Gamma\left(K_{3,k}-K_{4.l}\right)}.
\end{equation}
\hrulefill
\end{figure*}

\section{ Eavesdropping Attack on Both FSO and RF Links Simultaneously}
\subsection{Secrecy Outage Probability \textup{(SOP)}}

For a DF based FSO/RF system, the instantaneous secrecy capacity  can be obtained as 
\begin{equation}\label{eq:17}
C_{Inst}=min\left(C_{SR},C_{RD}\right)<R_{s}.
\end{equation}
where, $C_{SR}$ and $C_{RD}$ are the instantaneous secrecy capacities of SR and RD links respectively. Now the SOP in the presence of two independent eavesdroppers can be computed as 
\begin{align}\label{eq:18}
SOP_2&=Pr\left\{ min\left(C_{SR},C_{RD}\right)<R_{s}\right\}\nonumber \\ 
&=1-Pr\left\{ C_{SR}\geq R_{s}\right\} Pr\left\{ C_{RD}\geq R_{s}\right\}. 
\end{align}

Further this can be reformulated as 

\begin{align}\label{eq:19}
SOP_2&=1-\left(1-\int_{0}^{\infty}F_{SR}\left(\Theta\gamma_{SE_{1}}+\Theta-1\right)f_{SE_{1}}\left(\gamma_{SE_{1}}\right)d\gamma_{SE_{1}}\right)\nonumber\\
&\times\left(1-\int_{0}^{\infty}F_{RD}\left(\Theta\gamma_{RE_{2}}+\Theta-1\right)f_{RE_{2}}\left(\gamma_{RE_{2}}\right)d\gamma_{RE_{2}}\right)
\end{align}

As deriving the closed form of this SOP is very difficult, therefore we can derive the lower bound as follows

\begin{align}\label{eq:20}
SOP_{L_2}&\cong I_{1}\times I_{2}
\end{align}
with
\begin{align}\label{eq:21}
I_{1}=1-\left(1-\int_{0}^{\infty}F_{SR}\left(\Theta\gamma_{SE_{1}}\right)f_{SE_{1}}\left(\gamma_{SE_{1}}\right)d\gamma_{SE_{1}}\right)
\end{align}
and 
\begin{align}\label{eq:22}
I_{2}=\left(1-\int_{0}^{\infty}F_{RD}\left(\Theta\gamma_{RE_{2}}\right)f_{RE_{2}}\left(\gamma_{RE_{2}}\right)d\gamma_{RE_{2}}\right)
\end{align}

Using \eqref{eq:1} and \eqref{eq:2} in \eqref{eq:21} and \eqref{eq:3} and \eqref{eq:4} in \eqref{eq:22} and thereafter utilizing (12), the closed form expression for $SOP_{L}$ is obtained as \eqref{eq:23}, where
$$\varLambda=\sum_{k=1}^{m-1}\frac{1}{k!}\left(\frac{m\Theta}{\overline{\gamma}_{RD}}\right)^{k}\left(\frac{m}{\overline{\gamma}_{RE_{2}}}\right)^{m}\frac{1}{\Gamma\left(m\right)}\frac{\Gamma\left(m+k\right)}{\left(\frac{m\Theta}{\overline{\gamma}_{RD}}+\frac{m}{\overline{\gamma}_{RE_{2}}}\right)^{m+k}}$$ and $$\varpi=\left(\frac{\xi^{2}A}{2^{r}(2\pi)^{r-1}}\right)^{2}\sum_{m_{1}=1}^{\beta_{1}}b_{m_{1}}r^{\alpha+m_{1}-1}\sum_{m_{2}=1}^{\beta_{2}}b_{m_{2}}r^{\alpha+m_{2}-1}$$
\newline 
\emph{Proof}: See appendix B.
\newline
\begin{figure*}[t]
\begin{align}\label{eq:23}
SOP_{L_2}=\left[1-\left(1-\left(\frac{A\xi^{2}}{2^{r}(2\pi)^{r-1}}\right)^{2}\varpi G_{4r+3,4r+1,}^{3r+1,3r}\left[\frac{\overline{\gamma}_{SR}}{\Theta\overline{\gamma}_{SE_{1}}}\Bigg|\begin{array}{c}
1-K_{2},1,K_{1}\\
K_{2},0,1-K_{1}
\end{array}\right]\right)\times\varLambda\right]
\end{align}
\hrulefill
\end{figure*}

Moreover using [15, Eq. (41)], the $SOP_{L}$ can be expressed asymptotically at high SNR, as \eqref{eq:24}.
with $K_{i,j}$ denotes the $j_{th}$ term of $K_i$ and $i\in\left\{ 3,4\right\} $ with  $K_{3}=[1-\triangle\left(r,\xi^{2}\right),1-\triangle\left(r,\alpha\right),1-\triangle\left(r,m_{1}\right),1,\triangle\left(r,1+\xi^{2}\right)]$ and $K_{4}=[\triangle\left(r,\xi^{2}\right),\triangle\left(r,\alpha\right),\triangle\left(r,m_{2}\right),0,1-\triangle\left(r,2-\xi^{2}\right)]$.
\begin{figure*}[t]
\begin{align}\label{eq:24}
SOP_{L_2}^{\infty}\underset{\overline{\gamma}>>1}{\approxeq}1-\left\{ \left(1-\varpi\sum_{k=1}^{3r}\left(\frac{\overline{\gamma}}{\overline{\gamma}_{SE_{1}}\varTheta}\right)^{K_{3,k}-1}\frac{\prod_{l=1;l\neq k}^{3r}\Gamma\left(K_{3,k}-K_{3,l}\right)}{\prod_{l=3r+1}^{4r+1}\Gamma\left(1+K_{3,l}-K_{4.k}\right)}\times\frac{\prod_{l=1}^{3r+1}\Gamma\left(1+K_{4,l}-K_{3.k}\right)}{\prod_{l=3r+2}^{4r+1}\Gamma\left(K_{3,k}-K_{4.l}\right)}\right)\varLambda\right\} 
\end{align}
\hrulefill
\end{figure*}

\section{Strictly positive secrecy capacity analysis}
SPSC shows the probability of existence of secrecy capacity. It can be formulated as follows \cite{2}
\begin{equation}
SPSC_2=Pr\left\{ C_{s}\left(\gamma_{eq},\gamma_{SE}\right)>0\right\}\nonumber
\end{equation}
\begin{equation}\label{eq:25}
=1-SOP_{L_2}|_{R_{S}=0}.
\end{equation}
Using \eqref{eq:23} in \eqref{eq:25}, we obtain the SPSC as \eqref{eq:26}, where $$\varLambda^{'}=\sum_{k=1}^{m-1}\frac{1}{k!}\left(\frac{m}{\overline{\gamma}_{R_{1}D}}\right)^{k}\left(\frac{m}{\overline{\gamma}_{R_{1}E_{3}}}\right)^{m}\frac{\Gamma\left(m+k\right)}{\Gamma\left(m\right)\left(\frac{m}{\overline{\gamma}_{R_{1}D}}+\frac{m}{\overline{\gamma}_{R_{1}E_{3}}}\right)^{m+k}}$$

\begin{figure*}
\begin{equation}\label{eq:26}
SPSC_2=\left[\left(1-\left(\frac{A\xi^{2}}{2^{r}(2\pi)^{r-1}}\right)^{2}\varpi G_{4r+3,4r+1,}^{3r+1,3r}\left[\frac{\overline{\gamma}_{SR}}{\overline{\gamma}_{SE}}\Bigg|\begin{array}{c}
1-K_{2},1,K_{1}\\
K_{2},0,1-K_{1}
\end{array}\right]\right)\varLambda^{'}\right]
\end{equation}
\hrulefill
\end{figure*}

 The SPSC in Eq. \eqref{eq:26} can be obtained at high SNR as 
\begin{equation}\label{eq:27}
SPSC_{2}^{\infty}\underset{\overline{\gamma}>>1}{\approxeq}1-SOP_{L}^{\infty}|_{R_{S}=0}.
\end{equation}
Thus using \eqref{eq:24} in \eqref{eq:27}, we obtained the closed form expansion for SPSC at high SNR as \eqref{eq:28}

\begin{figure*}
\begin{equation}\label{eq:28}
SPSC_{2}^{\infty}\underset{\overline{\gamma}>>1}{\approxeq}\left\{ \left(1-\varpi\sum_{k=1}^{3r}\left(\frac{\overline{\gamma}}{\overline{\gamma}_{SE_{1}}}\right)^{K_{3,k}-1}\frac{\prod_{l=1;l\neq k}^{3r}\Gamma\left(K_{3,k}-K_{3,l}\right)}{\prod_{l=3r+1}^{4r+1}\Gamma\left(1+K_{3,l}-K_{4.k}\right)}\times\frac{\prod_{l=1}^{3r+1}\Gamma\left(1+K_{4,l}-K_{3.k}\right)}{\prod_{l=3r+2}^{4r+1}\Gamma\left(K_{3,k}-K_{4.l}\right)}\right)\varLambda^{'}\right\} 
\end{equation}
\hrulefill
\end{figure*}

 \section{Numerical Result} 
\begin{figure}[h!]
	\centering
\includegraphics[width=0.5\textwidth]{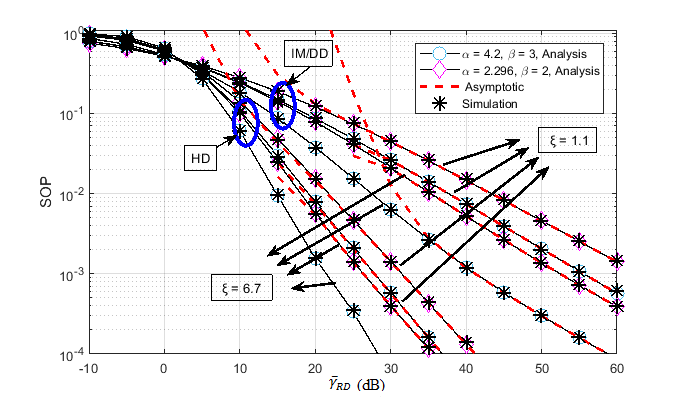}
	\caption{Secrecy outage probability versus average SNR with $m_{RD}$ = 2. }
	\label{fig:pic1}
\end{figure}
\begin{figure}[h!]
	\centering
\includegraphics[width=0.5\textwidth]{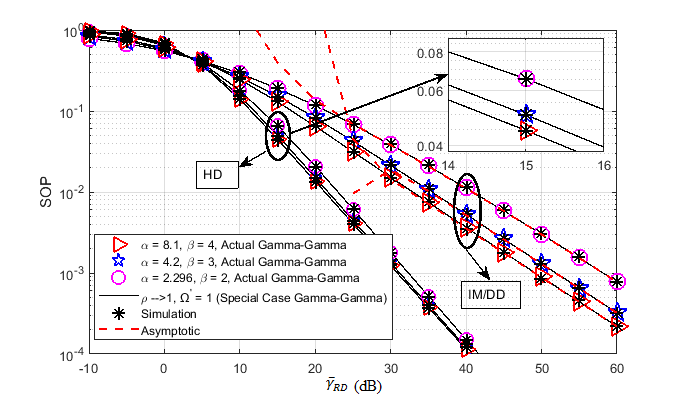}
	\caption{Secrecy outage probability versus average SNR for Gamma-Gamma special case with $m_{RD}$ = 1 and $\xi$ = 1.1. }
	\label{fig:pic1}
\end{figure} 
 
 \begin{figure}[h!]
	\centering
\includegraphics[width=0.5\textwidth]{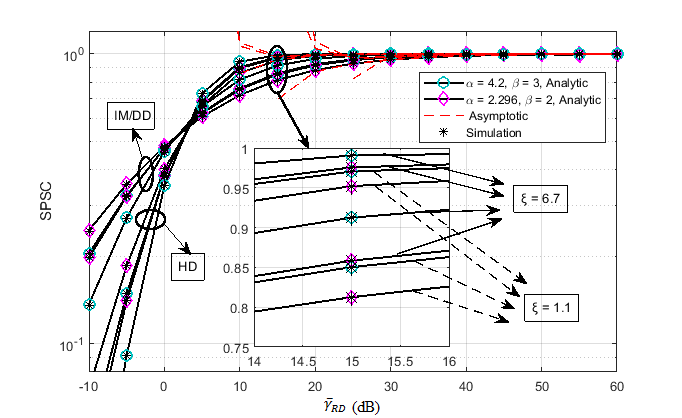}
	\caption{Strictly positive secrecy capacity versus average SNR with $m_{RD}$ = 2.}
	\label{fig:pic1}
\end{figure}
\begin{figure}[h!]
	\centering
\includegraphics[width=0.5\textwidth]{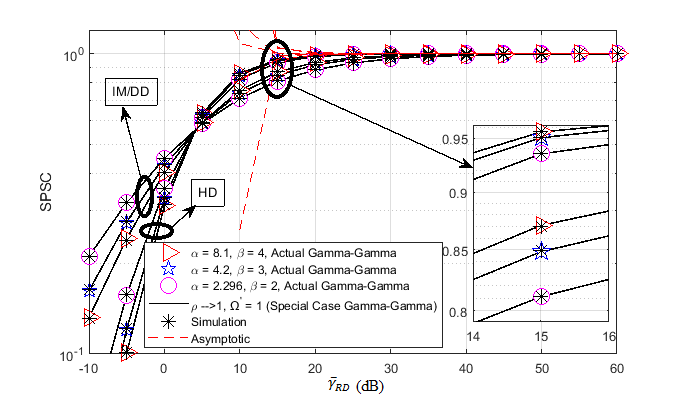}
	\caption{Strictly positive secrecy capacity versus average SNR for Gamma-Gamma special case with $m_{RD}$ = 1 and $\xi$ = 1.1. }
	\label{fig:pic1}
\end{figure}

\begin{figure}[t!]
	\centering
\includegraphics[width=0.5\textwidth]{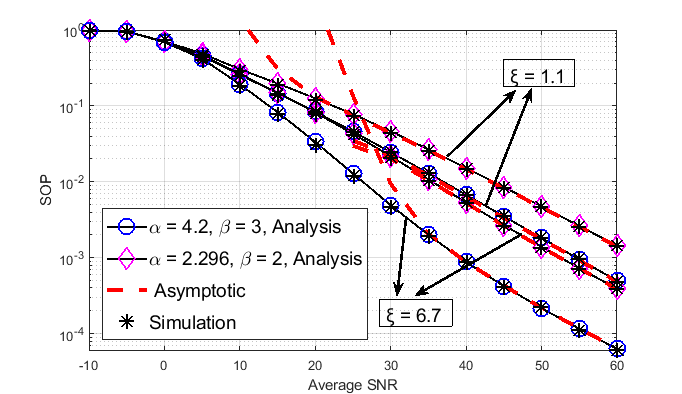}
	\caption{Secrecy outage probability versus average SNR (${\overline{\gamma}_{RD}} (dB)$) with $m_{RD}$ = 2  and $r$ = 2 in the presence of two Eavesdroppers.}
	\label{fig:pic1} 
\end{figure}
\begin{figure}[t!]
	\centering
\includegraphics[width=0.5\textwidth]{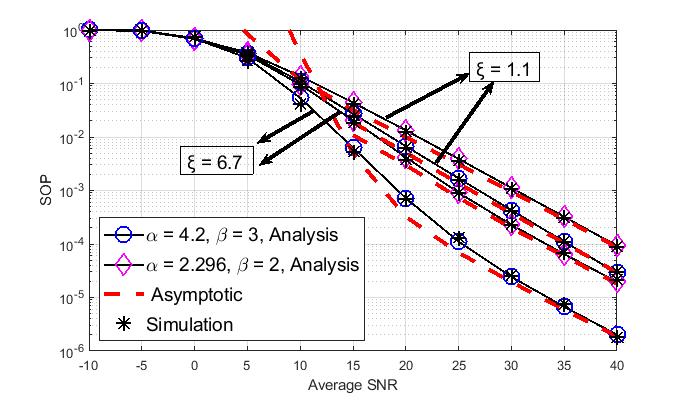}
	\caption{Secrecy outage probability versus average SNR (${\overline{\gamma}_{RD}} (dB)$) with $m_{RD}$ = 2  and $r$ = 1 in the presence of two Eavesdroppers. }
	\label{fig:pic1}
\end{figure}

 In this part, the analytical and asymptotic  results are presented to analyze the secrecy performance metrics. Furthermore, we analyze the impact of pointing deflection, type of detection and, fading parameters of FSO and RF links on SOP and SPSC. In this analysis, the secrecy rate is set to 0.01 nats/sec/Hz and  $m_1=m_2=m$. The turbulence parameters are considered as  $\alpha=2.296$ and $\beta=2$  for strong turbulence region and, for moderate turbulence region the values are $\alpha=4.2$ and $\beta=3$.

Fig. 2 shows the SOP versus ${\overline{\gamma}_{RD}}$ for different values of $r$, $\xi$, $\alpha$ and $\beta$. During analysis, the system is considered to be attacked by FSO based eavesdropper in the first hop. This figure reveals that the secrecy outage probability is improved with increasing in ${\overline{\gamma}_{RD}}$. It also concludes that the heterodyne detection scheme leads better secrecy performance than IM/DD scheme for different values of $\xi$. Moreover, the high pointing accuracy (i.e. high value of $\xi$) leads to better secrecy performance. From turbulence severity point of view, larger value of $\alpha$ and $\beta$ leads to better performance (i.e. smaller SOP), since the severity of the atmospheric turbulence gets lower. Fig. 3 demonstrates the effect of turbulence severity and detection techniques on SOP for a high deflection of pointing error. This figure is analysed for  Gamma-Gamma special case. This figure also results same as that for previous case (i.e. Fig. 2). 

Fig. 4 and Fig. 5 show the SPSC versus ${\overline{\gamma}_{RD}}$ with same parameter values used in the SOP analysis. The analysis is also done for different detection schemes with wide range of turbulence severity. These results reveal that the SPSC gets better with increasing ${\overline{\gamma}_{RD}}$. From Fig. 4, it is also observed that high pointing accuracy and large value of $\alpha$ and $\beta$ leads better result for SPSC.

Fig. 6 and Fig. 7  depict the SOP as a function of $R-D$ link's average SNR for $r = 2$ and $r = 1$. The curves are plotted by considering simultaneous attack of two eavesdroppers in both hops. All of these figures reveal that 
 HD ($r = 1$)detection scheme provides better result to that of IM/DD ($r = 2$). From turbulence severity point of view, it concludes that larger value of $\alpha$ and $\beta$ leads to better performance (i.e. smaller SOP), since the severity of the atmospheric turbulence gets lower. As shown in figures, Pointing deflection ($\xi$) plays a major role on the system performance. The low value of $\xi$ leads to worst secrecy performance as there is a chance of optical power received by the eavesdropper situated near the legitimate node. 

Finally, for the results presented in Fig. 1 -- Fig. 5, the asymptotic results tightly approximate with the closed form results. In the end, all the results are simulated through Monte-Carlo method.

 \section{Conclusion}
In this work, the analysis is done for two different scenarios such as: primarily it is assumed that an eavesdropper is attacking to FSO link only and secondly it is assumed that both RF as well as FSO links are attacked by two eavesdroppers simultaneously. The closed form expressions for SOP and SPSC are derived in terms of bivariate Fox's H-function and Meijer's G function. The impact of major parameters on the overall secrecy performance are depicted. In addition, The results show that low pointing error leads to  better result for the considered system. The detection method also influences the performance i.e. HD scheme leads to better performance as compared to the IM/DD scheme.

\vspace{0.15pc}
\appendices
  \section{Secrecy Outage Probability}
  \label{Secrecy Outage Probability}
\setcounter{equation}{0}
\renewcommand\theequation{\thesection.\arabic{equation}}

 Upon substituting Eq. \eqref{eq:1} and Eq. \eqref{eq:6} in Eq. \eqref{eq:8}, the integral is obtained as follows
\begin{multline}
SOP_{L}=\int_{0}^{\infty}\Bigg(1-\exp\left(-\frac{\varTheta\gamma}{\overline{\gamma}_{RD}}\right)\frac{A\xi^{2}}{2^{r}(2\pi)^{r-1}}\\\times\sum_{m_{1}=1}^{\beta}b_{m_{1}}r^{\alpha+m_{1}-1}G_{r+1,3r+1,}^{3r+1,0}\left[\frac{B^{r}\gamma}{\overline{r^{2r}\gamma}_{SR}}\Bigg|\begin{array}{c}
1,K_{1}\\
K_{2},0
\end{array}\right]\Bigg)\\\times\frac{A\xi^{2}}{2^{r}\gamma}\sum_{m_{2}=1}^{\beta}b_{m_{2}}G_{1,3}^{3,0}\left[B\left(\frac{\gamma}{\overline{\gamma}_{SE}}\right)^{\frac{1}{r}}\Bigg|\begin{array}{c}
\xi^{2}+1\\
\xi^{2},\alpha,m_{2}
\end{array}\right]d\gamma.
\end{multline}

After mathematical simplification, Eq. (A.1) can further be expanded as 
\begin{multline}
SOP_{L}=1-\frac{A\xi^{2}}{2^{r}(2\pi)^{r-1}}\frac{A\xi^{2}}{2^{r}\gamma}\sum_{m_{1}=1}^{\beta}b_{m_{1}}r^{\alpha+m_{1}-1}\sum_{m_{2}=1}^{\beta}b_{m_{2}}\\\times\int_{0}^{\infty}\gamma^{-1}\exp\left(-\frac{\varTheta\gamma}{\overline{\gamma}_{RD}}\right)G_{r+1,3r+1}^{3r+1,0}\left[\frac{B^{r}\gamma}{{r^{2r}\overline{\gamma}_{SR}}}\Bigg|\begin{array}{c}
K_{1},1\\
K_{2},0
\end{array}\right]\\\times G_{1,3}^{3,0}\left[B\left(\frac{\gamma}{\overline{\gamma}_{SE}}\right)^{\frac{1}{r}}\Bigg|\begin{array}{c}
\xi^{2}+1\\
\xi^{2},a,b
\end{array}\right]d\gamma.
\end{multline}

By expressing $exp (.)$ and $G_{p,q}^{m,n}\left[.\right]$ functions in their corresponding H-function representations using the identity [17, Eq. (07.34.26.0008.01)], the integral in terms of Fox's H-functions is obtained as  
\begin{multline}
SOP_{L}=1-\frac{A\xi^{2}}{2^{r}(2\pi)^{r-1}}\frac{A\xi^{2}}{2^{r}\gamma}\sum_{m_{1}=1}^{\beta}b_{m_1}r^{\alpha+m_1}\sum_{m_2=1}^{\beta}b_{m_2}\int_{0}^{\infty}\gamma^{-1} \\\times H_{0,1}^{1,0}\left[\frac{\varTheta\gamma}{\overline{\gamma}_{RD}}\Bigg|\begin{array}{c}
-\\
\left(0,1\right)
\end{array}\right] H_{r+1,3r+1}^{3r+1,0}\left[\frac{B^{r}\gamma}{r^{2r}\overline{\gamma}_{SR}}\Bigg|\begin{array}{c}
\left(K_{1},\left[1\right]_{k_{1}}\right),\left(1,1\right)\\
\left(K_{2},\left[1\right]_{k_{2}}\right),\left(0,1\right)
\end{array}\right]\\\times H_{1,3}^{3,0}\left[B^{r}\left(\frac{\gamma}{\overline{\gamma}_{SE}}\right)\Bigg|\begin{array}{c}
(\xi^{2}+1,r)\\
(\xi^{2},r),(a,r),(b,r)
\end{array}\right]d\gamma.
\end{multline}
 Now, using the formula [15, Eq. (2.3)], Eq. (A.3) can be expressed as Eq. \eqref{eq:9}.

  \section{Derivation of $I_{1}$ and $I_{2}$}
  \label{Derivation of $I_{1}$ and $I_{2}$}
  
\setcounter{equation}{0}
\renewcommand\theequation{\thesection.\arabic{equation}}

In this appendix, we derive $I_{1}$ and $I_{2}$ as follows:
\begin{equation}
I_{1}=1-\int_{0}^{\infty}F_{SR}\left(\Theta\gamma_{SE_{1}}\right)f_{SE_{1}}\left(\gamma_{SE_{1}}\right)d\gamma_{SE_{1}}
\end{equation}
Herein we only solved the integral part of $I_{1}$. Thus using \eqref{eq:1} and \eqref{eq:2} in the above integral, the obtained expression is written as 
\begin{multline}
I_{1}=1-\varpi\int_{0}^{\infty}G_{r+1,3r+1,}^{3r,1}\left[\frac{E\gamma}{\mu_{SR}}|\begin{array}{c}
1,K_{1}\\
K_{2},0
\end{array}\right]\\
\times G_{1,3,}^{3,0}\left[B\left(\frac{\gamma}{\mu_{SR}}\right)^{\frac{1}{r}}\Bigg|\begin{array}{c}
\xi^{2}+1\\
\xi^{2},\alpha,m_{2}
\end{array}\right]d\gamma
\end{multline}
By utilizing [ 17, Eq.(07.34.21.0013.01)] in the above integral, we obtain the closed form solution for $I_{1}$ as
\begin{equation}
I_{1}=1-\left(\frac{A\xi^{2}}{2^{r}(2\pi)^{r-1}}\right)^{2}\varpi G_{4r+3,4r+1,}^{3r+1,3r}\left[\frac{\overline{\gamma}_{SR}}{\Theta\overline{\gamma}_{SE_{1}}}\Bigg|\begin{array}{c}
1-K_{2},1,K_{1}\\
K_{2},0,1-K_{1}
\end{array}\right]
\end{equation}

Similarly, for $I_2$ can be formulated as
\begin{equation}
I_{2}=\left(1-\int_{0}^{\infty}F_{RD}\left(\Theta\gamma_{RE_{2}}\right)f_{RE_{2}}\left(\gamma_{RE_{2}}\right)d\gamma_{RE_{2}}\right)
\end{equation}
 
 Utilizing \eqref{eq:3} and \eqref{eq:4}, the integral part of the can be written as 
\begin{multline}
I_{2}=1-\int_{0}^{\infty}\left\{ 1-\sum_{k=0}^{m-1}\frac{1}{k!}\left(\frac{m\Theta\gamma}{\overline{\gamma}_{RD}}\right)^{k}\exp\left(-\frac{m\Theta\gamma}{\overline{\gamma}_{RD}}\right)\right\}\nonumber\\ \times\left(\frac{m}{\overline{\gamma}_{RE}}\right)^{m}\frac{\gamma^{m-1}}{\Gamma\left(m\right)}\exp\left(-\frac{m\gamma}{\overline{\gamma}_{RE}}\right)d\gamma
\end{multline} 
  Thus utilizing $\int_{0}^{\infty}x^{\alpha-1}\exp\left(-\mu x\right)dx=\frac{\Gamma\left(\alpha\right)}{\mu^{\alpha}}$ in the above expression, the solved closed for expression is obtained as 
 \begin{equation}
 I_{2}=\sum_{k=1}^{m-1}\frac{1}{k!}\left(\frac{m\Theta}{\overline{\gamma}_{RD}}\right)^{k}\left(\frac{m}{\overline{\gamma}_{RE}}\right)^{m}\frac{1}{\Gamma\left(m\right)}\frac{\Gamma\left(m+k\right)}{\left(\frac{m\Theta}{\overline{\gamma}_{RD}}+\frac{m}{\overline{\gamma}_{RE}}\right)^{m+k}}
\end{equation}

\end{document}